\documentclass[prd,showpacs,preprintnumbers,twocolumn,amsmath,nofootinbib,amssymb]{revtex4}
\usepackage{graphicx,color,dcolumn,booktabs,bm}
\usepackage{longtable,lscape}
\usepackage{txfonts}
\usepackage{overpic}
\usepackage{amssymb}
\usepackage{epstopdf}
\usepackage{indentfirst}
\usepackage{feynmf}   
\usepackage{slashed}  
\usepackage{cases}
\usepackage{color}
\usepackage{float}
\usepackage{multirow}
\usepackage{graphicx,color,dcolumn,booktabs,bm}
\usepackage{epsfig,dsfont,amssymb,amsmath,amsfonts,amsbsy,mathrsfs}

\graphicspath{{Figures/}} %
\usepackage[colorlinks, citecolor=blue,anchorcolor=red,menucolor=red, linkcolor=red,filecolor=red,runcolor=red,urlcolor=blue,frenchlinks=red]{hyperref}

\makeatletter
\@addtoreset{equation}{section}
\makeatother

\allowdisplaybreaks
\begin{document}
\title{Prediction for $5^{++}$ mesons}

\author{Cheng-Qun Pang$^{1}\footnote{Corresponding author}$}
\email{xuehua45@163.com}
\author{Ya-Rong Wang$^{1}$}
\author{Chao-Hui Wang$^{1}$}
\affiliation{$^1$College of Physics and Electronic Information Engineering, Qinghai Normal University, Xining 810000, China\\}
\begin{abstract}
In this paper, we study the spectrum and decay behavior of $5^{++}$ meson family which is still missing in experiment. By the {modified} Godfrey-Isgur model with a color screening effect, we obtain the mass spectrum of $a_5$, $f_5$ and $f_5^\prime$ mesons. And we predict their two-body strong decays by means of a phenomenology quark pair creation model. This study is crucial to { establish $J^{PC}=5^{++}$ meson family and it is also helpful to search for these states in the future}.
\end{abstract}

\maketitle

\section{introduction}\label{sec1}

{In the past decades, { the} quark model has made great achievements. There are six quarks, up($u$), down($d$), strange($s$), charm($c$), bottom($b$), and top($t$), in {the} quark model. Except $t$ quark, the other quarks can form mesons, baryons {,and other hadrons}. As an important part of hadron, {the} meson family is {phenomenologically} studied by many works} \cite{Wang:2012wa,Piotrowska:2017rgt,He:2013ttg,Chen:2015iqa,Ye:2012gu,Pang:2014laa, Wang:2014sea,Pang:2017dlw,Pang:2015eha}.
When checking the
{experimental} status of mesons \cite{Patrignani:2016xqp}, we notice an interesting phenomenon,
{the} whole meson family with $J^{PC}=5^{++}$ (with $q \bar{q}$ component, here $q$ defines $u$, $d$, or $s$ quark) is still missing, yet the other families for $a$ and $f$ meson (such as $a_0$, $a_1$, $a_2$, $a_3$, $a_4$, and $a_6$) have been reported by Particle Data Group(PDG) \cite{Patrignani:2016xqp}. This phenomenon stimulates our interest in exploring where the  mesons with $J^{PC}=5^{++}$ are and their properties.

{ The mesons with $J^{PC}=5^{++}$ have two families, which are isovector $a_5$ meson family, and isoscalar meson family{[}($f_5$($n\bar n$, here $n$  defines $u$ or $d$ quark) and $f_5^\prime$($s\bar s$)){]}.
Due to the present experimental progress on mesons, it is a suitable time to systematically carry out phenomenological study of missing $5^{++}$ states. This study is not only crucial to { establish $J^{PC}=5^{++}$   meson family and  helpful to search for these states in the future, but also important for verifying quark model.
\par
As $5^{++}$  mesons have a higher spin, the screening effect will be strong for the large angular momentum and larger average distance between quark pair.
 So we need introduce the screening effect into
the quark model in this work when we deal with the spectrum.
In this work, we calculate the mass spectra of the  $5^{++}$  meson family by using the unquenched Godfrey-Isgur(GI) model \cite{Song:2015nia,Song:2015fha},
which contains the screening effect.
  According to the former studies \cite{Godfrey:1985xj,Barnes:2005pb,Sun:2014wea,Song:2015nia,Song:2015fha,Godfrey:2015dia,Godfrey:2016nwn,Pang:2017dlw}, the GI model was tested by different systems, which shows that the GI model works well for describing hadron spectroscopy. In this work, we continue to apply this model to explore high spin states, especially estimate the mass spectrum of high spin mesons. Since the experimental information of high spin states is absent, we hope that these predicted states can be accessible at future experiment, which can provide further test of the GI model to high spin states. In this paper, we fix the parameters in the model by fitting some {well-established} meson states, which are adopted when calculating the mass of $5^{++}$  meson states. Then, for further study the properties of   $5^{++}$ mesons, we study their Okubo-Zweig-Iizuka (OZI)-allowed two-body strong decays taking input with the spatial wave functions obtaining in mass spectrum numerically calculation. Their partial and total decay widths are predicted by using a quark pair creation (QPC) model which was proposed by Micu \cite{Micu:1968mk} and extensively applied to studies of strong decay of other hadrons \cite{LeYaouanc:1972ae,
vanBeveren:1979bd,vanBeveren:1982qb,LeYaouanc:1988fx,roberts,Capstick:1993kb,Blundell:1995ev,Ackleh:1996yt,Capstick:1996ib,Bonnaz:2001aj,
Close:2005se,Zhang:2006yj,Lu:2006ry,Sun:2009tg,Liu:2009fe,Sun:2010pg,Rijken:2010zza,Yu:2011ta,Zhou:2011sp,Ye:2012gu,Wang:2012wa,
Sun:2013qca,He:2013ttg,Sun:2014wea,Pang:2014laa,Wang:2014sea,Chen:2015iqa}. We hope that our effort will be helpful to establish $a_5$, $f_5$ and $f_5^\prime$ meson families.}

This paper is organized as follows. In Sec. \ref{sec2}, the mass spectrum analysis of the meson family with $J^{PC}=5^{++}$ will be performed. In Sec. \ref{sec3}, we further study the two-body OZI-allowed strong decay behavior of these discussed states. The paper ends with a  conclusion in Sec. \ref{sec4}.

\section{THE MASS SPECTRUM ANALYSIS}\label{sec2}

In this work, the modified GI quark model is utilized to calculate the mass spectrum and wave functions of the  meson family.
In the following, this model will { be} illustrated in {detail}.

\subsection{The modified GI model}

In 1985, Godfrey and {Isgur} raise the GI model for describing relativistic meson spectra with great success, exactly in low-lying mesons\cite{Godfrey:1985xj}. As for the excited states, the screening potential must be taken into account for coupled-channel effect.

The interaction between quark and antiquark is depicted by the Hamiltonian of potential model including kinetic energy pieces and effective potential piece,
\begin{equation}\label{Hamtn}
  \tilde{H}=\sqrt{m_1^2+\mathbf{p}^2}+\sqrt{m_2^2+\mathbf{p}^2}+\tilde{V}_{\mathrm{eff}}(\mathbf{p,r}),
\end{equation}
where $m_1$ and $m_2$ denote the mass of quark and antiquark respectively, and effective potential $\tilde{V}_{\mathrm{eff}}$ contains two ingredients, a short-range $\gamma^{\mu}\otimes\gamma_{\mu}$ one-gluon-exchange interaction and a $1\otimes1$ linear confinement interaction. The meaning of tilde will { be explained} later.

In the nonrelativistic limit, effective potential has { a} familiar format\cite{Godfrey:1985xj,Lucha:1991vn}
\begin{eqnarray}
V_{\mathrm{eff}}(r)=H^{\mathrm{conf}}+H^{\mathrm{hyp}}+H^{\mathrm{so}}\label{1},
\end{eqnarray}
with
\begin{align}
 H^{\mathrm{conf}}&=\Big[-\frac{3}{4}(c+br)+\frac{\alpha_s(r)}{r}\Big](\bm{F}_1\cdot\bm{F}_2),\nonumber\\ &=S(r)+G(r)\label{3}\\
H^{\mathrm{hyp}}&=-\frac{\alpha_s(r)}{m_{1}m_{2}}\Bigg[\frac{8\pi}{3}\bm{S}_1\cdot\bm{S}_2\delta^3 (\bm r) +\frac{1}{r^3}\Big(\frac{3\bm{S}_1\cdot\bm r \bm{S}_2\cdot\bm r}{r^2} \nonumber  \\ \label{3.1}
&\quad -\bm{S}_1\cdot\bm{S}_2\Big)\Bigg] (\bm{F}_1\cdot\bm{F}_2),  \\
H^{\mathrm{so}}=&H^{\mathrm{so(cm)}}+H^{\mathrm{so(tp)}},  \label{3.2}
\end{align}
where $H^{\mathrm{conf}}$ includes the spin-independent linear confinement piece $S(r)$ and Coulomb-like potential from one-gluon-exchange $G(r)$, $H^{\mathrm{hyp}}$ denotes the color-hyperfine interaction consists tensor and contact terms, and $H^{\mathrm{SO}}$ is the spin-orbit interaction with
\begin{eqnarray}
H^{\mathrm{so(cm)}}=\frac{-\alpha_s(r)}{r^3}\left(\frac{1}{m_{1}}+\frac{1}{m_{2}}\right)\left(\frac{\bm{S}_1}{m_{1}}+\frac{\bm{S}_2}{m_{2}}\right)
\cdot
\bm{L}(\bm{F}_1\cdot\bm{F}_2),
\end{eqnarray}
colour magnetic term causing of one-gluon-exchange and
\begin{eqnarray}
H^{\mathrm{so(tp)}}=-\frac{1}{2r}\frac{\partial H^{\mathrm{conf}}}{\partial
r}\Bigg(\frac{\bm{S}_1}{m^2_{1}}+\frac{\bm{S}_2}{m^2_{2}}\Bigg)\cdot \bm{L},
\end{eqnarray}
the Thomas precession term.
\par
    For above formulas, $\bm{S}_1/\bm{S}_2$ indicates the spin of quark/antiquark and $\bm{L}$ the orbital momentum between them. $\bm{F}$ is relevant to the Gell-Mann matrix, i.e., $\bm{F}_1=\bm{\lambda}_1/2$ and $\bm{F}_1=-\bm{\lambda}^*_2/2$, and for a meson, $\langle\bm{F}_1\cdot\bm{F}_2 \rangle=-4/3$.
\par
     Now relativistic effects of distinguish influence must be considered especially in meson system, which is embedded in two ways. {First}, based on the nonlocal interactions and new $\mathbf{r}$ dependence, a smearing {function} is introduced for a meson $q\bar{q}$
\begin{equation}
\rho \left(\mathbf{r}-\mathbf{r'}\right)=\frac{\sigma^3}{\pi ^{3/2}}e^{-\sigma^2\left(\mathbf{r}-\mathbf{r'}\right)^2},
\end{equation}
which is applied to $S(r)$ and $G(r)$ to obtain smeared potentials $\tilde{S}(r)$ and $\tilde{G}(r)$ by
\begin{equation}\label{smear}
\tilde{f}(r)=\int d^3r'\rho(\mathbf{r}-\mathbf{r'})f(r'),
\end{equation}
with
\begin{eqnarray}
   \sigma_{12}^2=\sigma_0^2\Bigg[\frac{1}{2}+\frac{1}{2}\left(\frac{4m_1m_2}{(m_1+m_2)^2}\right)^4\Bigg]+
  s^2\left(\frac{2m_1m_2}{m_1+m_2}\right)^2,
\end{eqnarray}
where {the values of $\sigma_0$ and $s$ are defined later}.

{Second}, owning to relativistic effects, a general potential should rely on the mass-of-center of interacting quarks. Momentum-dependent factors{,} which will be unity in the nonrelativistic limit{,} are applied as
\begin{equation}
\tilde{G}(r)\to \left(1+\frac{p^2}{E_1E_2}\right)^{1/2}\tilde{G}(r)\left(1 +\frac{p^2}{E_1E_2}\right)^{1/2},
\end{equation}
and
\begin{equation}
  \frac{\tilde{V}_i(r)}{m_1m_2}\to \left(\frac{m_1m_2}{E_1E_2}\right)^{1/2+\epsilon_i} \frac{\tilde{V}_i(r)}{m_1 m_2} \left( \frac{m_1 m_2}{E_1 E_2}\right)^{1/2+\epsilon_i},
\end{equation}
where $\tilde{V}_i(r)$ {delegates} the contact, tensor, vector spin-orbit and scalar spin-orbit terms, and $\epsilon_i$ the relevant modification parameters.

Diagonalizing and solving the Hamiltonian in Eq.(\ref{Hamtn}) by exploiting a simple harmonic oscillator (SHO) basis, we will obtain the mass spectrum and wave functions.
{ As we know, a series of SHO wave function with different radial quantum number $n$ can be regarded as a complete basis to expand the exact radial wave function of meson state.
The base can produce all kind of states, even spurious ones(if the motion of the meson is taken into account \cite{Wang:1987pp}), However, this model is calculated in center-mass frame, so there are no spurious states in our numerical result.}

In configuration and momentum space, SHO wave functions have explicit {forms respectively},
\begin{align}
\Psi_{nLM_L}(\mathbf{r})=R_{nL}(r, \beta)Y_{LM_L}(\Omega_r),\nonumber\\
\Psi_{nLM_L}(\mathbf{p})=R_{nL}(p, \beta)Y_{LM_L}(\Omega_p),
\end{align}
with
\begin{eqnarray}
&R_{nL}(r,\beta)=\beta^{3/2}\sqrt{\frac{2n!}{\Gamma(n+L+3/2)}}(\beta r)^{L}
e^{\frac{-r^2 \beta^2}{2}} \nonumber \\
 &\times L_{n}^{L+1/2}(\beta^2r^2),\\
 &R_{nL}(p,\beta)=\frac{(-1)^n(-i)^L}{ \beta ^{3/2}}e^{-\frac{p^2}{2 \beta ^2}}\sqrt{\frac{2n!}{\Gamma(n+L+3/2)}}{(\frac{p}{\beta})}^{L} \nonumber \\
 &\times L_{n}^{L+1/2}(\frac{p^2}{ \beta ^2}),
\end{eqnarray}
where $Y_{LM_L}(\mathrm{\Omega})$ is spherical harmonic function, and $L_{n-1}^{L+1/2}(x)$ is the associated Laguerre polynomial, and $\beta=0.4~\mathrm{GeV}$ for our calculation.

{
The space-spin wave function $R_{nL}(r, \beta)\phi_{LSJM}$ with total angular quantum number $\it J$ can be constructed by coupling $L\otimes S$
\begin{align}
\phi_{LSJM}=\sum_{M_LM_S}C(L M_L S M_S; JM)Y_{LM_L}(\Omega_r)\chi_{SM_S},
\end{align}
here $C(L M_L S M_S; JM)$ is Clebsch-Gordan coefficient. {For the matrix element $\langle \alpha | \hat{V}(r,\hat{p}) | \beta \rangle$ where $| \alpha \rangle$ and $| \beta \rangle$ are arbitrary SHO basis with quantum number $\{n,J,L,S\}$ and $\{n^{\prime},J^{\prime},L^{\prime},S^{\prime}\}$}. The matrix element can be calculated conveniently by using SHO base as follows.
\begin{align}
&\langle \alpha | \hat{V}(r,\hat{p}) | \beta \rangle \nonumber \\
&={\langle \alpha | f(p)g(r) | \beta \rangle} \nonumber \\
&=\sum_{n}{\langle \alpha | f(p) |  n\rangle   \langle n |   g(r) | \beta \rangle}.
\end{align}
After diagonalizing the Hamiltonian matrix, we can obtain the mass and wave function of meson which are available to the following strong decay process.}

{ $5^{++}$ {mesons} have a higher spin, so the quarks and {antiquarks} will have large angular momentum and larger average distance which is greater than about 1 fm.
In this circumstance, light quark {antiquark} pairs will be spontaneously created and the screening effect will be strong.
 So we introduce the screening effect into
 GI model in this work when we deal with the spectrum.
 In the previous work \cite{Song:2015nia}, the modified GI model was proposed}, and the prediction results for the charm-strange mesons are consistent with the experimental data. For higher excitation states, the screen effect is considered to be very important by the authors of Ref. \cite{Song:2015nia} . It could be introduced by the transformation
 $br+c\rightarrow \frac{b(1-e^{-\mu r})}{\mu}+c$, {where $\mu$ is} screened parameter whose particular value is need to be fixed by the comparisons between theory and experiment. Modified confinement potential also requires similar relativistic correction, which has been mentioned in the GI model.
 Then, we further write $V^{\mathrm{scr}}(r)$ as the way given in Eq. (\ref{5}),
\begin{eqnarray}
\tilde V^{\mathrm{scr}}(r)&=& \int d^3 \bm{r}^\prime
\rho (\bm{r-r^\prime})\frac{b(1-e^{-\mu r'})}{\mu}.\label{5}
\end{eqnarray}
By inserting the form of  $\rho(\bm{r-r^\prime})$  in Eq.~(\ref{smear}) into the above expression and finishing this integration, the concrete expression for $\tilde V^{\mathrm{scr}}(r)$ is given by
\begin{eqnarray}
\tilde V^{\mathrm{scr}}(r)&=& \frac{b}{\mu r}\Bigg[r+e^{\frac{\mu^2}{4 \sigma^2}+\mu r}\frac{\mu+2r\sigma^2}{2\sigma^2}\Bigg(\frac{1}{\sqrt{\pi}}
\int_0^{\frac{\mu+2r\sigma^2}{2\sigma}}e^{-x^2}dx-\frac{1}{2}\Bigg) \nonumber\\
&&-e^{\frac{\mu^2}{4 \sigma^2}-\mu r}\frac{\mu-2r\sigma^2}{2\sigma^2}\Bigg(\frac{1}{\sqrt{\pi}}
\int_0^{\frac{\mu-2r\sigma^2}{2\sigma}}e^{-x^2}dx-\frac{1}{2}\Bigg)\Bigg]. \label{Eq:pot}
\end{eqnarray}
{Notably}, {except for}  converting the confinement potential to the screened potential, the other processing contents and the Hamiltonian matrix elements contained in the original GI model are calculated.

In our calculation, we need the spatial wave functions of the discussed  meson family with $J^{PC}=5^{++}$
which can be numerically obtained by the modified GI model.

\subsection{Mass spectrum analysis}
{GI model can describe the mass of ground states of the  mesons successfully, yet it does not  describe the excited states well}. Since unquenched effects are important for a heavy-light system, it is better to adopt the modified GI model (MGI) \cite{Song:2015nia,Song:2015fha} which uses a screening potential with a new parameter $\mu$. The parameter $\mu$ describes inverse of the size of screening.
In our previous work \cite{Pang:2017dlw}, we calculate the kaon family spectra  { by using} {the} MGI model. In this work we will use this MGI model to obtain the mass { spectrum} of  meson with $J^{PC}=5^{++}$.
 Beforehand, we need to adjust the parameters of MGI model by fitting with the {experimental} data. {So we fix the following twelve parameters listed in {Table}} \ref{SGIfit1}.
\renewcommand{\arraystretch}{1.2}
\begin{table}[htbp]
\caption{Parameters and their values in this work and GI models. \label{SGIfit1}}
\begin{center}
\begin{tabular}{cccc}
\toprule[1pt]\toprule[1pt]
Parameter &  This work &GI \cite{Godfrey:1985xj} \\
 \midrule[1pt]
$m_u$(GeV) &0.163&0.22\\
$m_d$(GeV) &0.163&0.22\\
$m_s$(GeV) &0.387&0.419\\
$b$(GeV$^2$) &0.221&0.18\\
$c$(GeV)&-0.240&-0.253\\
{$\sigma_0$(GeV)}&{1.799}&{1.80}\\
{$s$(GeV)}&{1.497}&{1.55}\\
$\mu$(GeV)&0.0635 &0 &  \\
$\epsilon_c$&-0.138& -0.168\\
$\epsilon_{sov}$&0.157&-0.035\\
$\epsilon_{sos}$&0.9726&0.055\\
$\epsilon_t$&0.893& 0.025\\
\bottomrule[1pt]\bottomrule[1pt]
\end{tabular}
\end{center}
\end{table}
{In {Table} \ref{fitmeason}, we select forty one experimental data of  meson listed in PDG
and optimize these meson masses to
determine {twelve} parameters in {Table} \ref{SGIfit1}.
This optimization has $\chi^2/n=82$ which is smaller than 2638 for the GI model as shown in {Table} \ref{fitmeason}.

\begin{table}[htbp]
\caption{The experimental data~\cite{Agashe:2014kda} fitted in our work. $\chi^2=\frac{({\rm {(Th-Exp)}/{Error}})^2}{D.O.F}$, where This work, Exp, and Error represent the theoretical value, experimental results, and experimental error, respectively.  We select some established { $q\bar{q}$ meson states} in PDG \cite{Agashe:2014kda} for our fitting.
The unit of the mass is GeV. \label{fitmeason}}
\begin{center}
\[\begin{array}{cccccccc}
\hline
\hline
 \text{Component}&\text{State} & \text{This work} & \text{GI\cite{Godfrey:1985xj}} & \text{Exp} & \text{error} \\
 \hline
\multirow{31}*{\it{n}$\bar{n}$}
 &1^1 \text{S}_0 & {0.1397} & 0.1524 & 0.13957 & 0.0001 \\
& 2^1 \text{S}_0 & {1.294 }& 1.293 & 1.3 & 0.1 \\
 &3^1 \text{S}_0 &{ 1.806 }& 1.874 & 1.812 & 0.012 \\
 &1^1 \text{P}_1 & {1.228 }& 1.219 & 1.2295 & 0.0032 \\
& 2^1 \text{P}_1 &{ 1.736} & 1.777 & 1.96 & 0.03 \\
 &3^1 \text{P}_1 & {2.117} & 2.236 & 2.24 & 0.035 \\
 &1^1 \text{D}_2 & {1.677} & 1.68 & 1.6722 & 0.00025 \\
& 2^1 \text{D}_2 &{ 2.056} & 2.135 & 2.005 & 0.025 \\
 &3^1 \text{D}_2 & {2.366} & 2.534 & 2.285 & 0.03 \\
 &1^1 \text{F}_3 & {2.002} & 2.033 & 2.032 & 0.04 \\
& 2^1 \text{F}_3 & {2.312} & 2.431 & 2.245 & 0.019 \\
& 1^3 \text{S}_1 & {0.7744} & 0.7713 & 0.77526 & 0.026 \\
 &2^3 \text{S}_1 & {1.424} & 1.456 & 1.465 & 0.04 \\
 &3^3 \text{S}_1 &{ 1.907} & 1.998 & 1.9 & 0.022 \\
& 4^3 \text{S}_1 & {2.258} & 2.435 & 2.265 & 0.0005 \\
 &1^3 \text{P}_0 & {1.191} & 1.087 & 1.474 & 0.016 \\
& 1^3 \text{P}_1 & {1.235} & 1.238 & 1.23 & 0.015 \\
 &2^3 \text{P}_1 &{ 1.756} & 1.818 & 1.647 & 0.032 \\
& 1^3 \text{P}_2 &{ 1.317} & 1.307 & 1.3183 & 0.012 \\
& 2^3 \text{P}_2 & {1.777} & 1.823 & 1.732 & 0.05 \\
 &1^3 \text{D}_1 & {1.646} & 1.664 & 1.72 & 0.04 \\
 &2^3 \text{D}_1 & {2.048} & 2.153 & 2.0 & 0.035 \\
 &3^3 \text{D}_1 & {2.364} & 2.557 & 2.265 & 0.02 \\
 &2^3 \text{D}_2 & {2.058} & 2.155 & 1.94 & 0.03 \\
 &3^3 \text{D}_2 & {2.370} & 2.553 & 2.225 & 0.04 \\
& 1^3 \text{D}_3 & {1.708} & 1.683 & 1.6888 & 0.0021 \\
 &2^3 \text{D}_3 & {2.074} & 2.131 & 1.982 & 0.014 \\
& 1^3 \text{F}_4 & {2.019} & 2.008 & 1.996 & 0.035 \\
 &2^3 \text{F}_4 & {2.322} & 2.407 & 2.237 & 0.01 \\
 &1^3 \text{G}_5 & {2.278} & 2.296 & 2.33 & 0.005 \\
 &1^3 \text{H}_6 & {2.501} & 2.558 & 2.45 & 0.13 \\
 \hline
 \multirow{11}*{{\it{n}}$\bar{s}$ or s$\bar{n}$}&1^1 \text{S}_0 & {0.4953} & 0.4625 & 0.4976 & 0.0004 \\
 &2^1 \text{S}_0 & {1.471} & 1.454 & 1.46 & 0.02 \\
 &1^3 \text{S}_1 & {0.916} & 0.9028 & 0.8958 & 0.0008 \\
 &2^3 \text{S}_1 & {1.567} & 1.579 & 1.414 & 0.015 \\
 &1^3 \text{P}_0 & {1.325} & 1.234 & 1.425 & 0.05 \\
 &1^3 \text{P}_2 & {1.450} & 1.428 & 1.4324 & 0.0013 \\
 &1^3 \text{D}_1 & {1.765} & 1.776 & 1.717 & 0.027 \\
 &1^3 \text{D}_3 & {1.826} & 1.794 & 1.776 & 0.007 \\
 &1^3 \text{F}_4 & {2.126} & 2.108 & 2.045 & 0.009 \\
 &1^3 \text{G}_5 & {2.378} & 2.388 & 2.382 & 0.024 \\
 \hline
 &\chi^2& {82} & 2638 &   &   \\
\hline
\hline
\end{array}\]
\end{center}
\end{table}
Of course, besides the mass spectrum mesons with $J^{PC}=5^{++}$ was calculated by the GI model, { Ref. \cite{Ebert:2009ub} also {gives} a  spectrum for $5^{++}$ meson and we will compare them later. Finally, we can obtain the mass spectrum of these four $5^{++}$ states by the MGI mode list in {Table} \ref{rho2omega2masstab} and compare our numerical result with GI model \cite{Godfrey:1985xj} and \cite{Ebert:2009ub}}.
\begin{table}[htbp]
\caption{The mass spectrum of $5^{++}$ states.
The unit of the mass is GeV. \label{rho2omega2masstab}}
\begin{center}
\[\begin{array}{cccccccc}
\hline
\hline
 \text{State} & \text{This work} & \text{GI\cite{Godfrey:1985xj}} &\text{Ebert\cite{Ebert:2009ub}} \\
 \hline
 a_5(1H) & {2.492} & 2.610& 2.359   \\
 a_5(2H) & {2.719} & 2.941 &-  \\
 a_5(3H) & {2.922} &3.255 &-  \\

 f_5(1H) & {2.492}& 2.610 & 2.359   \\
 f_5(2H) &{2.719} & 2.941 &-  \\
 f_5(3H)&{2.922} & 3.255 &-   \\

  f_5^\prime(1H) &{2.679} &2.771& 2.720   \\
 f_5^\prime (2H)&{2.914} &3.097 &-   \\
 f_5^\prime(3H) &{3.118} & 3.406&-     \\
\hline
\hline
\end{array}\]
\end{center}
\end{table}
\par
So we can conclude that
\begin{enumerate}

\item{The ground states of the $5^{++}$ states are still missing in experiments, and the predicted mass { is 2.492 GeV} for  $a_5/f_5 $, which are smaller than Ref. \cite{Godfrey:1985xj} and close to the result of {{Ref. \cite{Ebert:2009ub}}}. $f_5^\prime(1H)$ has the mass { of  2.771} GeV which is smaller than Ref. \cite{Godfrey:1985xj} and {Ref. \cite{Ebert:2009ub}}.}

\item{The first {excited} states of $a_5/f_5 $ and $f_5^\prime$ have the mass { of   2.719 GeV and 2.922}GeV, respectively. For the second {excited} states, $a_5/f_5 $(3H) and $f_5^\prime$(3H) have the mass { of 2.914 GeV and 3.118} GeV, respectively, which are also smaller than Ref. \cite{Godfrey:1985xj}.}

\end{enumerate}

The above conclusions are only from the point of mass spectra view and we will study their strong decays in the next section.

\section{THE DECAY BEHAVIOR ANALYSIS}\label{sec3}

\subsection{QPC {model}}

{The QPC model is used to obtain  Okubo-Zweig-Iizuka (OZI) allowed hadronic strong decays.
The QPC model is firstly proposed by Micu \cite{Micu:1968mk}}, which is further developed by Orsay group.\cite{LeYaouanc:1972ae,LeYaouanc:1973xz,LeYaouanc:1974mr,LeYaouanc:1977gm,LeYaouanc:1977ux}.
This model was widely applied to the OZI-allowed two-body strong decay of hadrons in Refs. \cite{vanBeveren:1979bd,vanBeveren:1982qb,Capstick:1993kb,Page:1995rh,Titov:1995si,Ackleh:1996yt,Blundell:1996as,
Bonnaz:2001aj,Zhou:2004mw,Lu:2006ry,Zhang:2006yj,Luo:2009wu,Sun:2009tg,Liu:2009fe,Sun:2010pg,Rijken:2010zza,Ye:2012gu,
Wang:2012wa,He:2013ttg,Sun:2013qca,Pang:2014laa,Wang:2014sea}.

For a decay process $A\to B+C$, we can write
\begin{eqnarray}
\langle BC|\mathcal{T}|A \rangle = \delta ^3(\mathbf{P}_B+\mathbf{P}_C)\mathcal{M}^{{M}_{J_{A}}M_{J_{B}}M_{J_{C}}},
\end{eqnarray}
where $\mathbf{P}_{B(C)}$ is a three-momentum of a meson $B(C)$ in the rest frame of a meson $A$. A superscript $M_{J_{i}}\, (i=A,B,C)$ denotes an orbital
magnetic momentum. The transition operator $\mathcal{T}$ is introduced to describe a quark-antiquark pair creation from vacuum, which has the quantum number
$J^{PC}=0^{++}$, i.e., $\mathcal{T}$ can be expressed as
\begin{eqnarray}
\mathcal{T}& = &-3\gamma \sum_{m}\langle 1m;1~-m|00\rangle\int d \mathbf{p}_3d\mathbf{p}_4\delta ^3 (\mathbf{p}_3+\mathbf{p}_4) \nonumber \\
 && \times \mathcal{Y}_{1m}\left(\frac{\textbf{p}_3-\mathbf{p}_4}{2}\right)\chi _{1,-m}^{34}\phi _{0}^{34}
\left(\omega_{0}^{34}\right)_{ij}b_{3i}^{\dag}(\mathbf{p}_3)d_{4j}^{\dag}(\mathbf{p}_4).
\end{eqnarray}
This is completely constructed in the form of a visual representation to reflect the creation of a quark-antiquark pair from vacuum, where the quark and antiquark are denoted by indices $3$ and $4$, respectively.

A dimensionless parameter $\gamma$ depicts the strength of the creation of $q\bar{q}$ from vacuum, where the concrete values of the parameter $R$ which will be discussed in the later section. $\mathcal{Y}_{\ell m}(\mathbf{p})={|\mathbf{p}|^{\ell}}Y_{\ell
m}(\mathbf{p})$ are the solid harmonics. $\chi$, $\phi$, and $\omega$ denote the spin, flavor, and color wave functions respectively, which can be treated separately.
Subindices $i$ and $j$ denote the color of a $q\bar{q}$ pair.

By the Jacob-Wick formula \cite{Jacob:1959at}, the decay amplitude is expressed as
\begin{eqnarray}
\mathcal{M}^{JL}(\mathbf{P})&=&\frac{\sqrt{4\pi(2L+1)}}{2J_A+1}\sum_{M_{J_B}M_{J_C}}\langle L0;JM_{J_A}|J_AM_{J_A}\rangle \nonumber \\
&&\times \langle J_BM_{J_B};J_CM_{J_C}|{J_A}M_{J_A}\rangle \mathcal{M}^{M_{J_{A}}M_{J_B}M_{J_C}},
\end{eqnarray}
and the general decay width reads
\begin{eqnarray}
\Gamma&=&\frac{\pi}{4} \frac{|\mathbf{P}|}{m_A^2}\sum_{J,L}|\mathcal{M}^{JL}(\mathbf{P})|^2,
\end{eqnarray}
where $m_{A}$ is the mass of an initial state $A$.
In our calculation, we need the spatial wave functions of the discussed  mesons. which can be numerically obtained by the modified GI model.

In the previous section, {we obtain the mass spectrum and wave functions of the mesons}. At the same time, we can use QPC model to study the strong decay of the $J^{PC}=5^{++}$ meson families by the means of these wave functions.

{As a phenomenological model of calculating strong decays of hadron, Quark pair creation (QPC) model was employed to
estimate the decay behaviors of hadron. However, QPC model cannot very precisely describe experimental data. To some extent, $\rho\to\pi\pi$ cannot be reproduced well as shown in {Table} \ref{decayfit}. In fact, it is a long-standing question not only for QPC model but also for other quark models like flux-tube model \cite{Wang:2017iai}. Just considering this point, usually we selected more typical channels to fix $\gamma$ value in QPC model, where a global fit is adopted. And then, this fixed $\gamma$ value is applied to calculate other decays.}
In this work we obtain $ \gamma=11.6$  by fitting the partial decay widths of 30 decay channels as shown in {Table} \ref{decayfit}.

\begin{table}[htb]
\centering%
\caption{The measured partial decay widths of 30 decay channels
and the comparison with theoretical calculation (the third
column). Here, the minimum of $\chi^2$ is $636$.   \label{decayfit}}
\begin{tabular}{lcccccccc}
\hline
\hline
   Decay channel&       Exp(MeV) \cite{Ye:2012gu,Pang:2017dlw} &      This work\\
\hline
$\rho\to \pi \pi$ & $151.2 \pm 1.2$     &68.9   \\ 
$b_{1}\rightarrow \omega\pi$&            142$\pm$8 & 191 \\ 
$\phi\rightarrow KK$&                2.08$\pm$0.02 &1.53   \\ 
$a_{2}\rightarrow \eta\pi$&            15.5$\pm$0.7 &2.65  \\ 
$a_2 \to \rho\pi$ & $75.0 \pm 4.5$& 66.2  \\ 
$a_2\rightarrow KK$&        5.2$\pm$0.2 & 4.53  \\ 
$\pi_2\rightarrow f_2(1270)\pi$&       145.8$\pm$5.1 &  48.6 \\ 
$\pi_2\rightarrow \rho\pi$&            80.3$\pm$2.8 &220  \\ 
$\pi_2 \to K^*{K}$ & $10.1 \pm 3.4$    &26.7 \\ 
$\rho_{3}\rightarrow \pi\pi$&           38$\pm$2.4 & 61.8  \\ 
$\rho_{3}\rightarrow \omega\pi$&        25.8$\pm$1.6 &44.3 \\ 
$\rho_{3}\rightarrow K\overline{K}$&    2.5$\pm$0.2 &2.43 \\ 
$f_2 \to \pi \pi$ & $156.8 \pm 3.2$     &136 \\ 
$f_2 \to K\bar{K}$ & $8.6 \pm 0.8 $ s    &8.31 \\ 
$f_4 (2050) \to \omega\omega$ & $54 \pm 13 $  & 70.1 \\ 
$f_4 (2050) \to \pi\pi$ & $35.4\pm 3.8$    &79 \\ 
$f_4 (2050) \to KK$ & $1.4 \pm 0.7$     &0.941 \\ 
$f_2'(1525) \to K\bar{K}$ & $61 \pm 5$    & 53.7 \\ 
$K^{*}(892)\rightarrow K\pi$&                48.7$\pm$0.8 &   18.6 \\ 
$K^{*}(1410)\rightarrow K\pi$&                15.3$\pm$1.4 &  62.2 \\ 
$K^{*}_{0}(1430)\rightarrow K\pi$&                251$\pm$74 &291  \\ 
$K^{*}_{2}(1430)\rightarrow K\pi$&             54.4$\pm$2.5 &  50.2  \\ 
$K^{*}_{2}(1430)\rightarrow K^{*}\pi$&    26.9$\pm$1.2 &  19.9  \\ 
$K^{*}_{2}(1430)\rightarrow K\rho$&             9.5$\pm$0.4&  7.18   \\ 
$K^{*}_{2}(1430)\rightarrow K\omega$&          3.16$\pm$0.15 &  2.13 \\ 
$K_3^*(1780)\to K \rho$ & $74 \pm 10$     &25.8 \\ 
$K_3^*(1780)\to K^*\pi$ & $45 \pm 7$     &28.3 \\ 
$K_3^*(1780)\to K\pi$ & $31.7\pm 3.7$     &38.1 \\ 
$K_4^*(2045)\to K\pi$ & $19.6\pm 3.8$    &22 \\ 
$K_4^*(2045)\to K^* \phi$ & $2.8 \pm 1.4$    &34.8  \\ 
\hline
$\gamma =11.6$ &    &$\chi^2=636$  \\ 
\hline
\hline
\end{tabular}
\end{table}
Next, we will {analyze} the strong decay {behavior} of these $5^{++}$ states.
\subsection{ The ground states}
     The ground state $a_5$ which is not observed in experiment is predicted in this work, with the mass { of 2492 MeV ($a_5(2492)$)}, and the total width is { 396} MeV. $\rho_3\pi$ is its dominant decay channel, the width is about { 137} MeV, and the branch ratio is $0.36$. $a_2\rho$, $\omega\rho$, $\rho\pi$ {,}and $h_1\rho$, are its important decay channels which have the branch ratio about 0.08 each one, just as shown in {Table} \ref{a51H}.  The final states $b_1\omega$, $f_2\pi$, $f_4(2050)\pi$, $a_1\rho$ and $\rho(1450)\pi$, also have sizable decay widths, in which $b_1\omega$ and $f_2\pi$ almost {have the same} width about 20 MeV.

     As the {isospin} partner of $a_5(1H)$, we predict $f_5(1H)$ will have the mass {of 2.49} GeV and { the} width {of 327} MeV, respectively. In the final decay channels of $f_5(1H)$, $\rho\rho${,} and $b_1\rho$ will be the most important final states which  have the {widths of} 73 MeV, and their branch {ratios} are about 0.22. $a_2\pi$ and  $a_1\pi$  are the important decay channels too, with the widths { of}  58 MeV and { 54} MeV, respectively. In addition, $h_1\omega$ and  $\pi_2\pi$ also have visible widths {of} { 30} MeV and { 28} MeV which { are} presented in {Table} \ref{a51H}.
     The widths of $KK_3(1780)$ and $\rho(1450)\rho$ are very small(see {Table } \ref{a51H}), their branch { ratios are} about { 0.017}.

      $f_5^\prime(1H)$ is  the $s\bar{s}$ partner of $f_5(1H)$, has the mass  { of 2.68} GeV and  { the}  width more than 850 MeV in our prediction. As shown in {Table} \ref{a51H}, $f_5^\prime(1H)$  {mainly decays to two strange mesons for its $s\bar{s}$  component}. $KK_3^*(1780)$ is the dominant decay mode with the width { 161} MeV.  $K^*K_2^*(1430)$,  $K_1K^*$  are also the important decay channels {whose}  widths are over 100 MeV. Besides, $K^*K^*$, $KK_2^*(1430)$, $KK^*(1410)$, $KK^*$ are its sizable final channels with the branch {ratios} about 0.08. $KK_1$, $KK_4^*(2045)$, $K^*K_1^\prime$, $KK^*(1680)$, $KK_0^*(1430)$ are the visible decay channels of  $f_5^\prime(1H)$. Here, we do not consider the mix of the flavor between $f_5(1H)$ and $f_5^\prime(1H)$.

   \begin{table}[htb]
\centering%
\caption{The partial decay widths of the ground states for $5^{++}$ family, the unit of  widths is MeV.   \label{a51H}}
\[\begin{array}{cccccc}
\hline
\hline
\multicolumn{2}{c}{{a _5\text{(1H)}}} &\multicolumn{2}{c}{f _5\text{(1H)}}&\multicolumn{2}{c}{f _5^\prime \text{(1H)}}\\%
\\
\midrule[1pt]
 Total&{396}&Total&{327}&Total&{851}\\
 \midrule[1pt]
Channel &Value &Channel &Value&Channel &Value    \\
 \rho_3\pi  &{137 } &\rho\rho&{72.7 }&KK_3^*(1780) &{ 161 } \\
a_2\rho  &{54.2}   & b_1\rho &{72.9}   &K^*K_2^*(1430)&{144} \\
\omega\rho  &{35.6} & a_2\pi    &{58.4}&K_1 K^* &{111}\\
\rho\pi  &{31.5}  & a_1\pi    &  {54.2}& K^*K^*   & {85.6} \\
h_1 \rho  &{ 31.0}&h_1\omega    &{30.3} & KK_2^*(1430)&{76.3} \\
b_1 \omega & {23.6}&\pi_2\pi & {27.6 }&KK^*(1410) &{73.0} \\
f_2\pi &{ 21.4 }    &  KK_3^*(1780) & {5.63}         &KK^*  &{60.9} \\
 f_4(2050)\pi & {19.0}& \rho(1450)\rho  &{5.38} &KK_1 &{48.9 }\\
  a_1 \rho  & {18.1} &&&KK_4^*(2045)&{49.4}\\
\rho(1450)\pi &{ 16.7}&&   &  K_1^\prime K^*  &{26.2} \\
 KK_3^*(1780)& {5.63}& &  &  KK^*(1680) &{13.4} \\
  \omega(1420)\rho&{2.05}&   & & f_1(1425)\eta^\prime &{1.31} \\%

 \hline
 \hline
\end{array}\]
\end{table}

\subsection{The first {excited}  states}
\begin{table}
\centering%
\caption{The partial decay widths of the first {excited}  states for $5^{++}$ family, the unit of  widths is MeV.   \label{a52H}}
\[\begin{array}{cccccc}
\hline
\hline
\multicolumn{2}{c}{{a _5\text{(2H)}}} &\multicolumn{2}{c}{f _5\text{(2H)}}&\multicolumn{2}{c}{f _5^\prime \text{(2H)}}\\%
\midrule[1pt]
Channel &Value &Channel &Value&Channel &Value    \\
Total &{159}&Total &{113}&Total &{637}    \\
\rho_3\pi  &{76.9}&\rho(1450)\rho &{43.5} &K^*K^*(1680) & {111}\\
   a_0(1450)\rho   &{ 17.9} & \omega(1420)\omega  &{18.0}  &KK_3^*(1780)&{96.0}\\
  \rho(1450)\pi&{16.9}&\rho \rho &{9.66}
  & KK^*(1680) &{ 54.5}
  \\
\pi_2\rho&{ 9.81 }& a_1 \pi   &{8.19}& KK^*&{ 51.6}  \\

 a_1b_1 &{ 8.97}& \omega_3\omega  & { 7.01 }& K^*K^*&{ 46.7} \\
  \omega_3\rho &{ 7.51} &a_2\pi & {6.91} & K^*K^*(1410)&{43.3} \\

\rho_3 \omega &{ 5.48} & a_1a_2 &{ 5.88}&  K_1 K_2^*(1430)  &{38.8}\\

\omega\rho& {4.76}&a_2a_2&{ 2.51}&K K_1 &{37.6} \\
 f_2a_1&{2.90}&\omega\omega &{2.34}&K^*K_3^*(1780)  &{28.1} \\
 f_2\pi&{2.82}&\pi_2\pi &{2.01}& K_1K^* &{ 27.8} \\
         a_2f_2 &{2.08}  &f_1f_2 &{ 1.75}&K K_2^*(1430)&{26.8}\\
a_2b_1&{1.72} &f_2f_2& { 1.63 }& K^*K_2^*(1430) &{22.7}\\
a_2f_1&{ 1.65 } & b_1b_1&{ 1.63 }   &K_1^\prime K^*(1410)&{16.7}  \\
 & &a_1a_1 & { 1.45 }&K_1K^*(1410)&{15.8}\\
& &&& K^*(1410)K_2^*(1430)&{7.32}  \\
&&  & & K_1K_1^\prime &{5.67}   \\
&&& &  KK_1^\prime &{3.97}  \\
  & & &&  KK^*(1410)&{2.72}\\
 \hline
 \hline
\end{array}\]
\end{table}
In this section,  we will {analyze} the strong decay behavior of the first {excited}  states of $5^{++}$ family.

$a_5(2H)$ has the mass {of 2719} MeV and narrow width {of} {159} MeV in our theory result. According to {Table } \ref{a52H}, $\rho_3\pi$ is the dominant decay channel of $a_5(2H)$ which is similar { to} $a_5(1H)$, the branch ratio is about 0.49. $\rho(1450)\pi$ and $a_0(1450)\rho$ are its important decay channels which have the branch ratios about 0.1, just as {Table} \ref{a52H}  shown.  The final decay modes $\pi_2\rho$, $\omega_3\rho$, $a_1b_1$, $\omega\rho$ and $\rho_3\omega$, also have sizable decay widths,  with the width {of}5-10} MeV. The other decay information is shown in {Table} \ref{a52H}.

  $f_5(2H)$ as the {isospin} partner(I=0) of $a_5(2H)$ has the  mass {of 2.72} GeV and width {of} 113 MeV, respectively. $f_5(2H)$ {mainly} decays to  $\rho(1450)\rho$ which has the  width {of} 44 MeV and the branch ratio 0.25.
    $\omega(1420)\omega$, $\rho\rho$ and $a_1\pi$ are the important final states which  have the { widths of 18 MeV, 9.7 MeV and 8.2 MeV}, respectively, and their branch {ratios} are $0.15$, $0.11$ {,}and $0.10$, respectively. $a_2\pi$,  $\omega_3\omega$ $a_1a_2${,} and $\omega\omega$ also have visible {widths} from 2.3 MeV to 7 MeV which are presented in {Table} \ref{a52H}. The widths of other channels
    are very small(see {Table} \ref{a52H}) {and their}  branch ratios are less than 0.02.

 As the $s\bar{s}$ partner of $f_5(2H)$, $f_5^\prime(2H)$  has the mass  {of 2.91} GeV and { the } width  {of 637} MeV in our prediction. According to Tab \ref{a52H}, $f_5^\prime(2H)$ also {mainly} decays into two kaon mesons. $K^*K^*(1680)$ and  $KK_3^*(1780)$ are the most important decay modes with the {widths of 111 MeV and 96 MeV}, respectively. $KK^*(1680)$, $KK^*$, $K^*K^*$ {and $K^*K^*(1410)$} are also the more important decay channels {whose widths are} in the range of {43-54} MeV. In addition, $K_1K_2^*(1430)$, $KK_1$,  $K^*K_3^*(1780)$, $K_1K^*$, $KK_2^*(1430)$ and $K^*K_2^*(1430)$,  are its sizable final channels {whose branch ratios}  in the range of {0.04-0.07}.
\subsection{ The second {excited}  states}
\begin{table}
\centering%
\caption{The partial decay widths of the second {excfited}  states for $5^{++}$ family, the unit of  widths is MeV.   \label{a53H}}
\[\begin{array}{cccccc}
\hline
\hline
\multicolumn{2}{c}{{a _5\text{(3H)}}} &\multicolumn{2}{c}{f _5\text{(3H)}}&\multicolumn{2}{c}{f _5^\prime \text{(3H)}}\\%
\midrule[1pt]
Channel &Value &Channel &Value&Channel &Value \\
Total &{103}&Total &{57.2}&Total &{522}  \\

 \rho_3\pi  &{37.7} &\rho\rho &{13.6}&K_1K_2^*(1430)&{101}\\

 \rho(1450)\pi  &{ 12.8}&    \rho(1450)\rho&{ 10.1} &K^*K^*(1410)& { 94.8}\\

 a_1b_1&  {6.97}  &   a_1a_2&{6.23}   &KK_3^*(1780) & {58.0}\\
  \omega\rho  &{6.74}   &\rho_3\rho &{5.8}

 &KK^*(1680)&{ 52.3}\\

    \rho\pi  &{6.68} &a_0(1450) a_1    & {5.1}&K^*K_3^*(1780)&{48.5} \\

\pi_2\rho    & {6.58}&\omega(1420)\omega&{4.48} &KK^*&{ 41.5}\\

\omega_3\rho    &{5.82}&a_2a_2 &{3.82} & K^*K^*(1680)&{30.9} \\

\omega(1420)\rho    &  {4.70}& \pi_2\pi    & {2.76}& KK_1&{20.4}\\

\omega\rho(1450) & { 3.23}&a_1 a_1   & { 2.76}&K_1K^*(1410)&{ 11.9}\\

a_2\rho    & { 2.92}&b_1b_1 & { 1.87}&K^*K_1&{11.3}\\

f_2    a_2& {2.57}&a_1 \pi    &  {0.674}& K_1^\prime K^*(1410)&{10.7} \\

a_2b_1& {2.14}&    & &K^*K_2^*(1430)&{8.55}\\

f_2    a_1& {2.90}&   &  &K^*(1410) K^*(1410) & {8.03} \\

b_1h_1&{ 1.43}&    & &KK^*(1410) & {7.72}\\

 & &    &  &K_1K_1&{7.37}\\

 & &    & &K_1K_1^\prime&{6.78} \\

& &    &&K_1K^*(1680)&{2.61}\\

 \hline
 \hline
\end{array}\]
\end{table}

 We also calculate the two body strong decays of the second {excited}  states of $5^{++}$ family.

 As the {isovector} meson of $5^{++}$ family, $a_5(3H)$  has the mass {of 2.92} GeV, the total width  {of} {103} MeV which is very narrow. $\rho_3\pi$ also is its dominant decay channel as shown in {Table} \ref{a53H}, the width is about 37.7 MeV, and the branch ratio is $0.4$. $\rho(1450)\pi$ has a large ratio (0.13) in its decay final channels. $\rho\pi$, $\omega\rho$, $a_1b_1$, $\pi_2\rho$, $\omega_3\rho$, $\omega(1420)\rho${,} and $\rho(1450)\omega$ also have sizable contribution in the total widths.

$f_5(3H)$ state is the second  radial excited state of $f_5$, with the mass of 2.92  GeV and width of 57 MeV. $f_5(3H)$  mainly  decays into $\rho\rho$ and $\rho(1450)\rho$, whose decay widths are  13.6 MeV and of 10.1  MeV, respectively, and each channel almost has the branch ratio 0.2. $a_1a_2$, $\rho_3\rho$, and $\omega(1420)\omega$ modes are the important decay channels too, with the widths about 5 MeV. In addition, $a_2a_2$, $a_0(1450)a_1$, and $\pi_2\pi$ also have visible widths which are presented in {Table} \ref{a53H}.
     The width of other modes are very small({see {Table} \ref{a53H}}).

      $f_5^\prime(3H)$ has the $s\bar{s}$ component as the partner of  $f_5(3H)$ which has the mass {of 3.12 GeV}. $f_5^\prime(3H)$ has the total width  {of 522 MeV} in our calculation. Just as shown in {Table} \ref{a53H}, $f_5^\prime(3H)$ { mainly decays} to $K_1K_2^*(1430)$ and $KK^*(1410)$ with the {widths of 101 MeV and 95 MeV}.
   $KK_3^*(1780)$, $KK^*(1680)$, $K^*K_3^*(1780)${,} and $KK^*$ are its important decay  channels  with the {widths of 58 MeV, 52 MeV, 49 MeV{,} and 42 MeV, respectively.}

   $K^*K^*(1680)$, $KK_1$, $K_1K^*(1410)$, $K^*K_1${,} and  $K_1^\prime K^*(1410)$
   are its sizable final channels  with the branch {ratios} of 0.06, 0.04, 0.023, 0.022{,} and 0.02, respectively.
   Besides,    $K^*K_2^*(1430)$, $K^*(1410)K^*(1410)$, $KK^*(1410)$, $K_1K_1${,} and $K_1K_1^\prime)$ have the visible contribution to the total width too. The other modes have very small { widths} in the final states of  $f_5^\prime(3H)$.
\section{CONCLUSION}\label{sec4}
In this paper, we study the spectrum and two body strong decay of the family with $J^{PC}=5^{++}$ which is still missing in experiment. By the modified Godfrey-Isgur model with a color screening effect, we {analyze} the mass spectrum of $a_5$ and $f_5$ mesons, in which we find that the ground states of the $5^{++}$ states,
$a_5$, $f_5${,} and $f_5^\prime$
 have the mass  {of 2.492 GeV, 2.492 GeV{,} and 2.68 GeV} and the {widths of 400 MeV, 330 MeV, and 850 MeV }, respectively.
The first {excited} states of $a_5/f_5 $ and $f_5^\prime$ have the mass {of 2.719 GeV and 2.922 GeV} and
$a_5(3H)/f_5 (3H)$ and $f_5^\prime(3H)$ have the mass  {of 2.914 GeV and 3.118  GeV, respectively}.
The total widths are predicted to be 160 MeV($a_5$(2H)), 110 MeV($f_5$(2H)), and 640 MeV($f_5^\prime$(2H)) for the first  {excited}  states. For the second {excited} states of $5^{++}$, $a_5$(3H), $f_5$ (3H){,} and $f_5^\prime(3H)$ have the widths  {of} {100} MeV($a_5$), {60} MeV($f_5$){,} and {430} MeV($f_5^\prime$), respectively.

We also predict the detail {of} decay information of $5^{++}$ family using QPC model which can be helpful {to search} the mesons in the future experiments just as BESIII and COMPASS.

\section{ACKNOWLEDGMENTS}
We thanks Xiang Liu, Jia-lun Ping{,} and Zhi-feng Sun for helpful communications and discussions.  This work is supported in part by {the Nature Science Foundation Projects of Qinghai Office of Science and Technology, No. 2017-ZJ-748, the Chunhui Plan of China's Ministry of Education, No. Z2017054.}
\bibliographystyle{apsrev4-1}
\bibliography{hepref}
\end{document}